# Structural, electrical and energy storage properties of lead-free NaNbO$_3$-BaHfO$_3$ thin films

Huijuan Dong[1], Bingcheng Luo[1a)], Kexin Jin[1]

[1]School of Physical Science and Technology, Northwestern Polytechnical University, Xi'an, Shaanxi, 710072, China

a) Electronic mail: luobingcheng@nwpu.edu.cn

**Abstract**

Lead-free dielectric thin-film capacitors with desirable energy storage density are gathering attention due to the increasing environmental concern and the integrating electronic devices. We here reported a series of new highly-orientated (1-$x$)NaNbO$_3$-$x$BaHfO$_3$ ($x$≤0.15) lead-free thin films prepared by a sol-gel method, and presented the dependence of their structural, electrical and energy storage properties on the $x$ level of BaHfO$_3$. The microstructure, leakage current and breakdown strength of pristine NaNbO$_3$ thin films are significantly improved by addition of BaHfO$_3$. As a result, the superior energy storage performances were obtained at $x$=0.1 with recoverable energy storage density $W_r$ of 23.1 J/cm$^3$ at 1100 kV/cm, excellent thermal stability from 30 to 210 °C ($\Delta W_r$ < 4.6%), good fatigue resistance ($\Delta W_r$ < 2.9% after 10$^4$ electrical cycles), and the fast charge-discharge rate ($\tau$ = 0.82 μs ).

## 1. Introduction

Dielectric capacitors, one of the typical solid-state energy storage components, have gained increasing attention due to their fast charging/discharging capability and large output power density for potential power electronics and pulse power applications [1]. However, their inferior energy density is still insufficient to compete with batteries and supercapacitors. Therefore, designing and exploring novel dielectric materials with satisfactory energy storage performance including high energy storage density and efficiency, excellent thermal stability and fatigue resistance, is one of the central tasks for dielectric research community. Generally, the energy storage density ($W_s$), the recoverable energy density ($W_r$) and the energy storage efficient ($\eta$) of dielectric capacitor can be evaluated by the integral of polarization-electric field (*P-E*) loops, *i.e.*, $W_s = \int_0^{P_m} EdP$, $W_r = \int_{P_r}^{P_m} EdP$, $\eta = W_r/W_s$, where $E$ is the applied electric field, $P_m$ and $P_r$ are the maximum polarization and remnant polarization, respectively [2]. Obviously, the high applied electric field $E$, large $P_m$ and low $P_r$ are highly desirable to achieve the superior energy-storage performance. Accordingly, antiferroelectric (AFE) dielectric with double *P-E* hysteresis loop (*i.e.*, $P_r$~0 and high $P_m$ in field-induced ferroelectric phase), is expected to be a competitive candidate for high energy-storage material.

NaNbO$_3$ (NNO), one of alkaline niobates, is a AFE oxide material. Historically, the primary concern in this material was on its complex structure and phase transitions. Several phase transitions in NNO were observed in the temperature regime of 20 to 1000K, including the ferroelectric phase (<153 K), the AFE P phase (153~633 K), the AFE R phase (633~753 K), and three other phase transitions in the temperature range of 753 to 1000 K [3]. This wide-temperature AFE characteristic makes NNO a promising lead-free alternative for lead-based energy storage materials due to the increasing health and environmental concerns, and thus a renewed enthusiasm is stimulated recently [4]. Their energy storage performances, however, often suffer from the field-induced metastable ferroelectric (FE) phase in NNO, which leads to the square-shaped rather than double *P-E* hysteresis loops observed experimentally. This is due to the similar free energies between coexisting AFE phase and field-induced metastable FE phase, and the latter one can remain after the removal of the applied field, leading to a large remanent polarization. Accordingly, destabilizing the induced FE phase in NNO through lowering the Goldschmidt tolerance factor is the primary strategy for energy storage engineering [5]. It was widely demonstrated that solid solution with certain perovskite compounds (*e.g.*, $CaZrO_3$, $SrTiO_3$, $Bi(Mg, Ta)O_3$ and others), could stabilize the AFE phase and/or induce the relaxor behavior in NNO bulk ceramics, resulting in the improved energy storage performances [6-12]. There are, however, limited reports for the energy storage properties of NNO thin films [13-15], which is preferred over bulk ceramics for potential applications in microelectronic devices, especially the portable electronic devices. In this work, we grew the $(1-x)NaNbO_3$-$xBaHfO_3$ ($x$=0, 0.05, 0.1, 0.15) thin films by a sol-gel method, and investigated their structural, electrical and energy storage properties. The cubic perovskite insulator $BaHfO_3$(BHO) was chosen here due to two primary reasons for improving the energy storage performance of NNO. One is the wide band gap of BHO ($E_g$=6.1 eV) [16], which could improve the breakdown strength and the thermal stability of NNO. The other is that Hf substitution in the Nb site could improve the polarizability and stabilize the AFE characteristic of NNO [5].

## 2. Experimental details

(1-$x$)NaNbO$_3$-$x$BaHfO$_3$ (NNO-$x$BHO, $x$=0, 0.05, 0.1, 0.15) thin films with thickness of ~200 nm were fabricated on(001) 0.7wt% Nb-doped SrTiO$_3$ (NSTO, MTI Corp.) single crystal substrates. Briefly, different compositions of NNO thin films were grown on NSTO substrates by the sol-gel method. The precursor solutions with concentration of 0.5 mol/L were used and spin-coated on substrates to obtain the thin films, which were annealed at the optimum temperature 680 °C for 10 min in a rapid thermal processor. The more details for fabrication process could be found elsewhere [13, 17].

X-ray diffraction (XRD, PANalytical Empyrean) and atomic force microscopy (AFM, Asylum Research MFP-3D) were used for the structural and morphology identification. For measuring the electrical characteristics, top Au circle electrodes with a diameter of 200 μm were sputtered on the film surface to form Au/NNO-$x$BHO/NSTO capacitors. Leakage current measurements were performed at room temperature by a Keithley 6517 meter. Dielectric characteristics were measured at room temperature by an Agilent E4980 LCR meter with a driving voltage of 50 mV. *P-E* hysteresis loops were analyzed by a radiant precision work-station at a frequency of 1 kHz.

## 3. Results and discussion

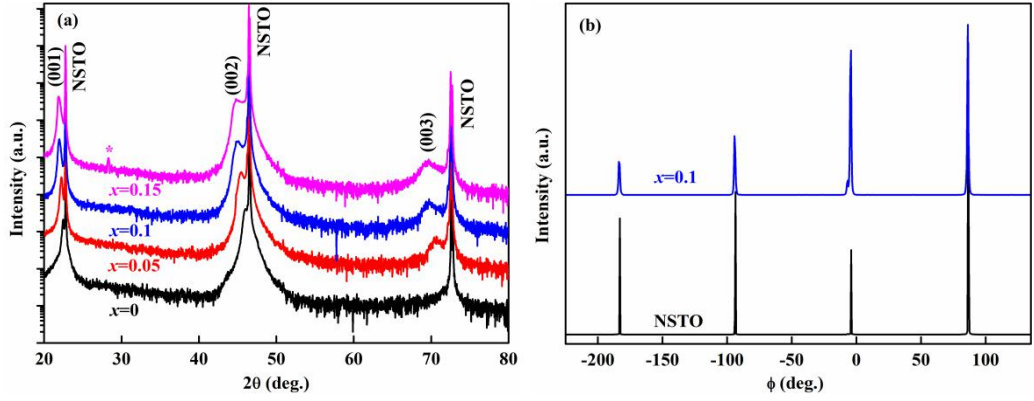

*Figure 1 (a) XRD patterns of NNO-xBHO thin films. Symbol (\*) represents the secondary phase. (b) Phi-scan patterns of x=0.1 thin films deposited on NSTO substrates*

Figure 1 (a) shows the XRD patterns of NNO-$x$BHO thin films deposited on NSTO substrates. All the peaks from NNO-$x$BHO thin films are near the corresponding peaks from NSTO substrates, demonstrating a perovskite structure with highly-preferred (001)-orientation. These peaks shift toward lower $2\theta$ angles with increasing the concentration of BHO, reflecting the expansion of the out-of-plane lattice constant. This is related to the larger ionic radii of $Ba^{2+}$ [0.161 nm, coordination number (CN)=12] and $Hf^{4+}$ [0.071 nm, CN=6], in comparison with host material NNO [$Na^+$: 0.139 nm CN=12; $Nb^{5+}$: 0.064 nm, CN=6] [18], wherein BHO is incorporated to the host lattice. A weak peak at $28.3^0$ (labeled as \*) could be seen in the $x$=0.15 thin film, implying a tiny secondary phase, which may be ascribed to the limitation of solid solubility of BHO in NNO. Additionally, the XRD Phi-scan measurements were performed to check the in-plane texture relationship between NNO-$x$BHO and NSTO. The typical patterns of the (202) peak for $x$=0.1 thin film are shown in Figure 1 (b). The four-fold symmetric peaks are seen for both $x$=0.1 thin film and NSTO substrate and their corresponding positions coincide well with each other, suggesting a cube-on-cube growth mode.

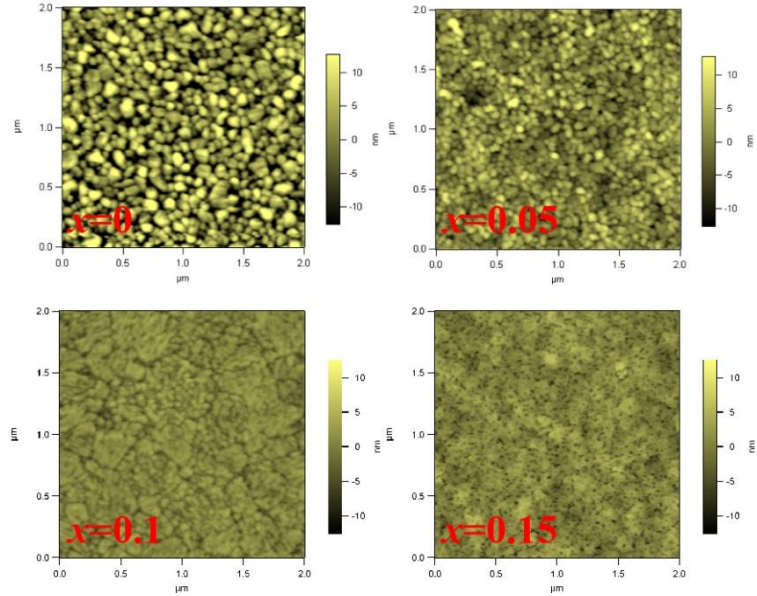

Figure 2 AFM images of NNO-xBHO thin films

Figure 2 presents the AFM images of NNO-$x$BHO thin films with 2μm×2μm scan size and the same scale. The surface morphologies are strongly related to the BHO content. The un-doped NNO thin film exhibits a rough surface, whereas the denser surfaces with smaller grains are seen in BHO-doped NNO thin films. Smaller grain size will be beneficial for energy storage purposes, because grain size is expected to be inversely associated with the breakdown strength of the dielectric materials [19]. The decreasing grain size with the incorporation of BHO into NNO thin films means the suppression of grain growth rate, commonly reflecting the reduced number of oxygen vacancies, as observed widely in chemical-solution-derived niobate-based thin films with aliovalent doping [13, 20, 21]. Additionally, the root-mean-square (RMS) of surface roughness determined through Igor Pro software is about 6.9±0.1 nm, 4.5±0.1 nm, 1.8±0.1 nm, and 1.9±0.1 nm for $x$=0, $x$=0.05, $x$=0.1, and $x$=0.15, respectively.

RMS values decrease systematically with the *x* content, with the exception of the *x*=0.15 thin film. This discrepancy in the *x*=0.15 thin film might be due to the influence of impurity phase as revealed in XRD.

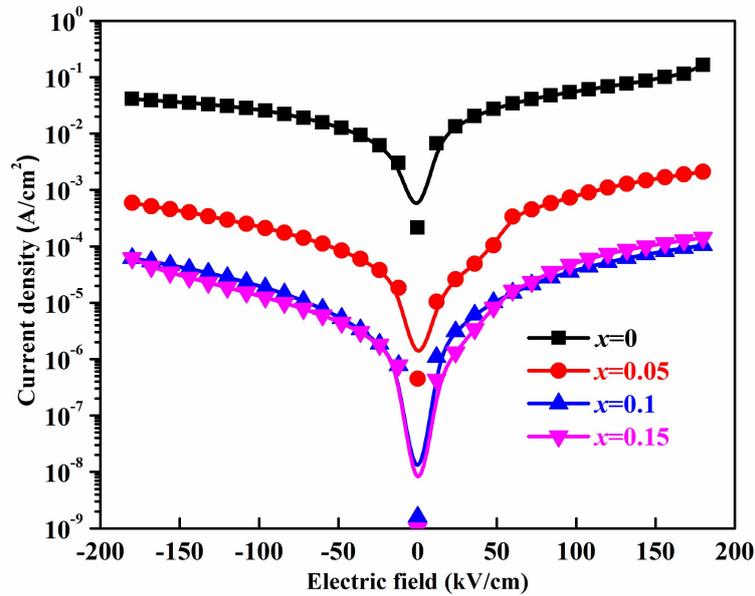

*Figure 3 Current density (J) vs. electric field (E) curves for NNO-xBHO thin films*

The leakage currents of NNO-*x*BHO thin films measured at room temperature are compared in Figure 3. The leakage current is reduced significantly in NNO-*x*BHO thin films from $6.05 \times 10^{-2}$ A/cm$^2$ (*x*=0) to below $6.02 \times 10^{-5}$ A/cm$^2$ (*x*=0.1 and *x*=0.15) at 120 kV/cm, which exhibits the lower value up to three orders of magnitude in *x*≥0.1 thin films. Additionally, the asymmetrical characteristics between negative and positive applied voltage are evident, mainly originating from the larger work functions of Au (~5.4 eV) [22] and NSTO (~4.2 eV) [23] in comparison with the electron affinity of NNO (3.53 eV) [24]. It is also seen that this asymmetry is more conspicuous in un-doped NNO thin film. Specifically, the current ratio at fixed field, *e.g.*, *I*(+180 kV/cm)/*I*(-180 kV/cm), decreases from 4.2 for *x*=0 to 3.5 for *x*=0.05 and 1.7 for *x*=0.1, and then increases up to 2.2 for *x*=0.15. Generally, the leakage behavior is vulnerable to several factors including microstructure and defects. For NNO thin film, sodium ions are easily volatilized during thermal processing. This volatility would introduce the cation-site vacancies and thus lead to the formation of oxygen vacancies due to the requirement of charge neutrality, which could provide the excess carriers resulting in the high leakage current. It is also noted that aliovalent doping in NNO thin films could suppress the formation of oxygen vacancies and improve the microstructure, both of which are in favor of lowering the leakage current [13, 21]. In our case, the more obvious asymmetry observed in *J-E* curves implies the existence of more defects in NNO-*x*BHO thin films, because defects could provide the excess carriers, leading to an upward shift of the Fermi level and the lowering barrier height between film and electrode. Thus, it is reasonably concluded that the improved microstructure and the reduced defects would be considered to be the main factors for the reduced leakage current in the present case.

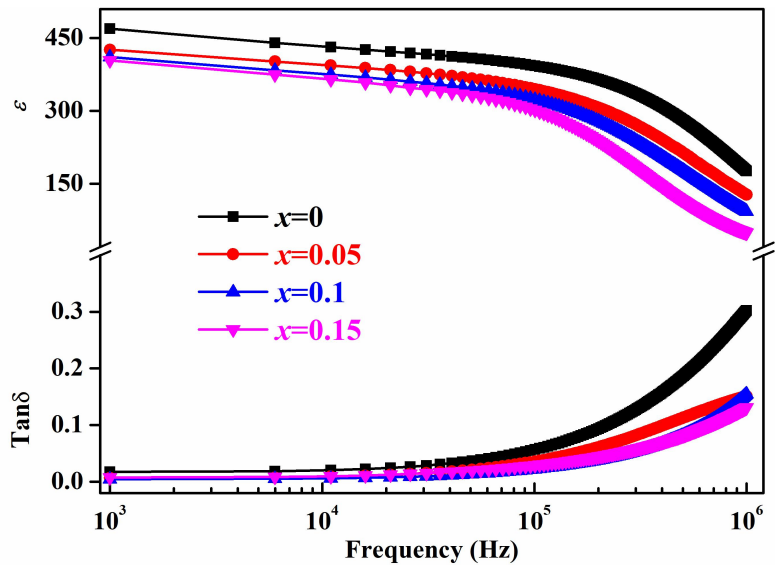

*Figure 4 Frequency-dependent dielectric constant* (a) *and dielectric loss* (b) *of NNO-xBHO thin films*

Figure 4 displays the frequency-dependence of dielectric constant ($\varepsilon$) and dielectric loss (*Tan δ*) for NNO-*x*BHO thin films measured at room temperature. First of all, the dielectric constant decreases with increasing the frequency, indicating the relaxation behavior, which is normally attributed to the reorientation of dipoles lagging behind the switching external field with the increment of frequency. It is also noted that the clear drop of dielectric constant within the higher frequency regime of >$10^5$ Hz is analogous to the behaviors observed in $BaTiO_3$-based and $BiFeO_3$-based thin films, which is commonly attributed to the influence of the space-charge polarization at the electrode-film interface [25, 26]. Secondly, the dielectric constant at the same frequency decreases systematically with the *x* content, for example, from 469.2 for *x*=0 to 408.1 for *x*=0.15 at 1 kHz. The decreased dielectric constant seems to be related to grain size revealed in Figure 2, *i.e.*, smaller grain size corresponding to lower dielectric constant, which is similar to the reported trend in dielectric thin films [27, 28]. It should be mentioned that the moderate dielectric constant and low dielectric loss are of benefit for a high breakdown strength [28]. Lastly, for the dielectric loss, it increases with the increment of frequency, which may stem from the LC resonance caused by the stray inductance of the contacts [29]. It is also noted that the dielectric loss is depressed with the addition of BHO. This decrease is mainly attributed to the reduced leakage current, which suppresses the conduction loss.

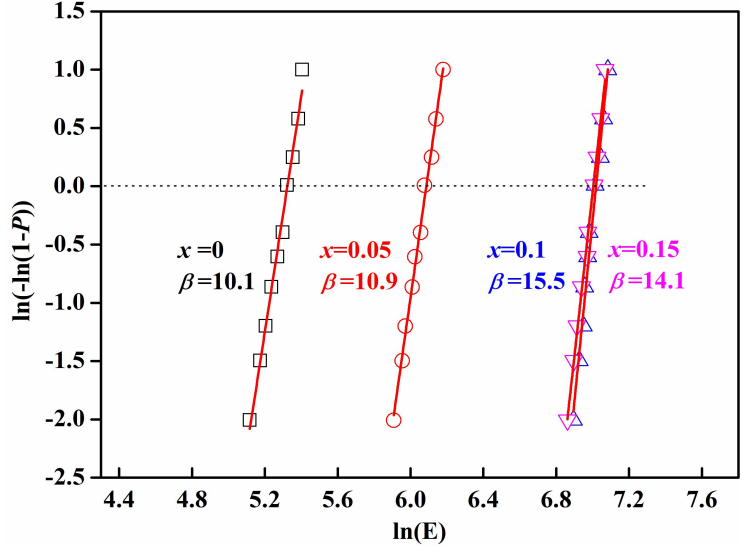

*Figure 5 The Weibull distribution and the corresponding fitting results for NNO-xBHO thin films*

For energy storage purpose, breakdown strength ($E_b$) is a very important parameter for dielectric capacitors, which can be estimated by the Weibull statistic distribution method, *i.e.*, $P(E)=1-\exp[-(E/E_b)^\beta]$, ($E$: the experimental breakdown strength, $P(E)$: the cumulative probability of dielectric breakdown, $E_b$: the characteristic breakdown strength at the probability of 63.2%, $\beta$: the shape parameter) [30]. As shown in Figure 5, the good linearity with high $\beta$ (>10) value is evident for all the samples. The $E_b$ value is calculated to be about 204.7 kV/cm, 435.9 kV/cm, 1112.1 kV/cm, and 1104.8 kV/cm for $x=0$, $x=0.05$, $x=0.1$, and $x=0.15$, respectively. The $E_b$ value is enhanced significantly with addition of BHO, which is closely related to the improved microstructure and electrical homogeneity mentioned above. Notably, in comparison with pristine NNO thin film, the maximum $E_b$ of over 4 times of magnitude is achieved at $x=0.1$, which will be beneficial for energy storage performance.

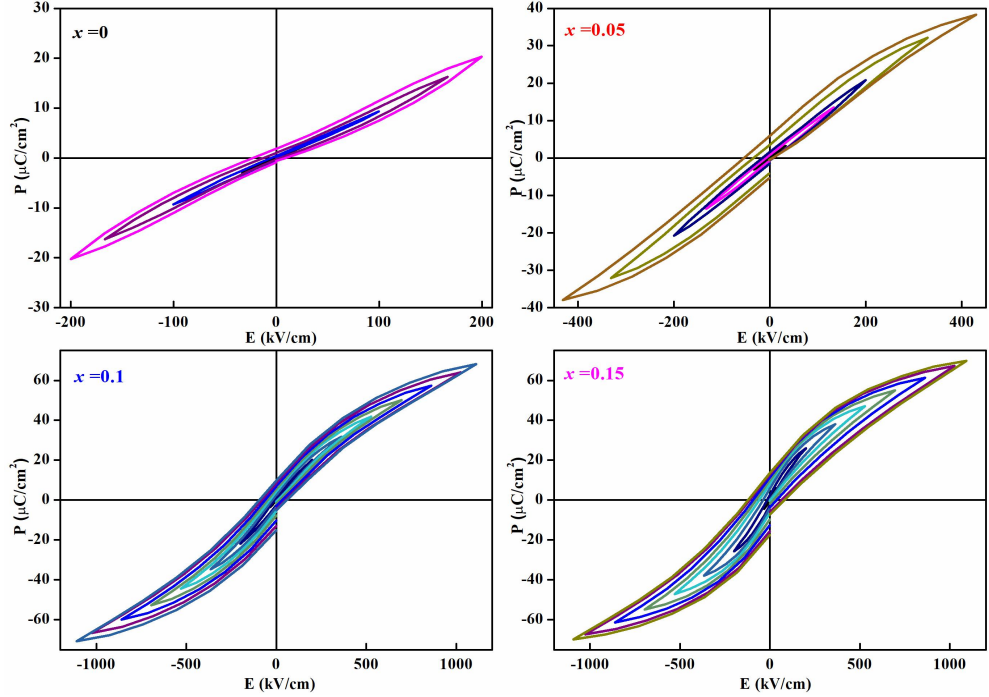

*Figure 6 P-E hysteresis loops for NNO-xBHO thin films under different electric fields*

With obtaining the $E_b$, the room-temperature *P-E* hysteresis loops for NNO-*x*BHO thin films were measured under different applied fields up to the proximity of their corresponding $E_b$ values, as shown in Figure 6. The pristine NNO thin film shows a pinched hysteresis loop without saturation polarization, whereas BHO containing thin films exhibit normal ferroelectric-like hysteresis loops with saturation polarization. This behavior means that introducing BHO into NNO may stabilize the field-induced FE phase due to the larger tolerance factor *t* of BHO (*t*=1) than that of NNO (*t*=0.967) [5, 31]. Further, we can extract field-dependent recoverable energy storage density $W_r$ and efficiency *η* from these *P-E* hysteresis loops, as illustrated in Figure 7. For all the thin films, $W_r$ increases with the increment of applied electric field and *η* shows the opposite trend. A maximum $W_r$ value of ~23.1 J/cm³ at 1100 kV/cm, along with an efficiency of 66.2 % is achieved in *x*=0.1 thin film, both of which are comparable to those of NNO-based thin films [13-15]. Additionally, $W_r$ is improved with addition of BHO. For example, even under low electric field, the $W_r$ values of $x$=0.1 thin films are still larger than those of pristine NNO thin film, as shown in inset of Figure 7. This observation infers that the superior energy storage properties in $x$=0.1 thin film are not only related to the external reasons including the improved microstructure and electrical homogeneity, but also associated with the intrinsic crystal chemistry. Further works will be needed.

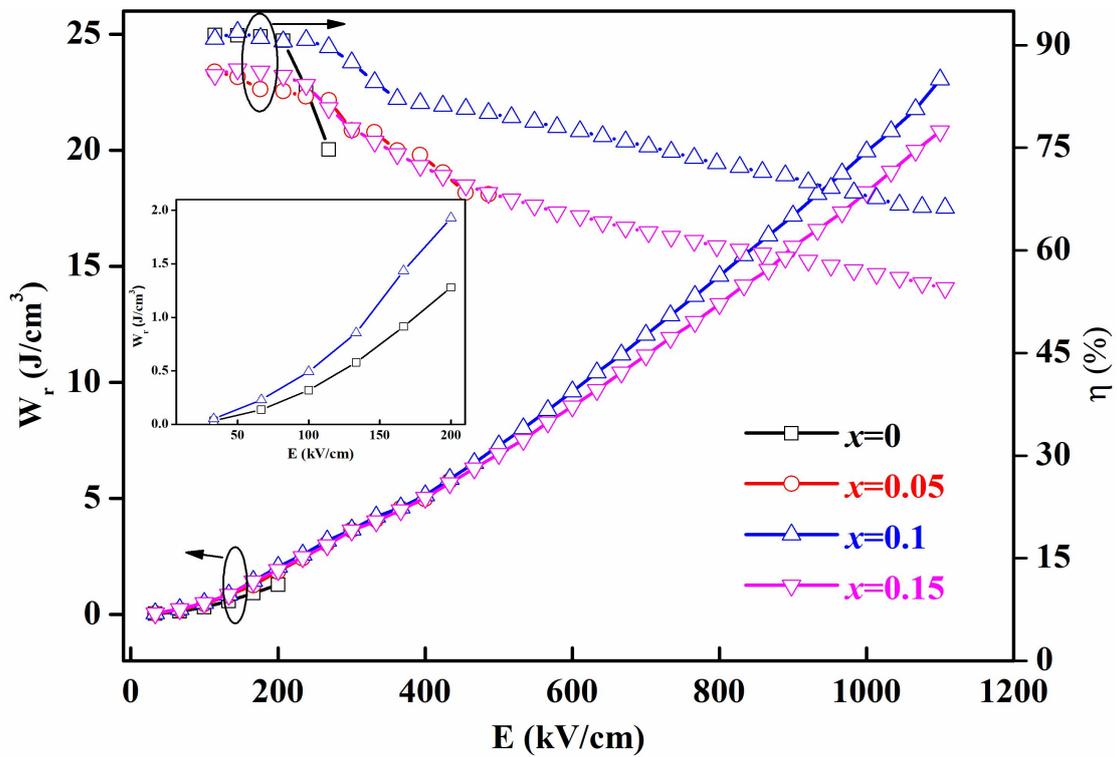

*Figure 7 $W_r$ and $\eta$ values of NNO-xBHO thin films as a function of electric fields. Inset shows an enlarged view of $W_r$ curves for x=0 and x=0.1 thin films*

Regarding the satisfactory energy storage density of $x$=0.1 thin film in NNO-$x$BHO, its thermal stability, fatigue resistance and charging-discharging performance were studied further, which is important for practical application. Temperature-dependent P-E hysteresis loops were measured firstly and the values of $W_r$ and $\eta$ were extracted, as shown in Figure 8 (a). The P-E hysteresis loops keep the similar shape within the temperature regime from 30 °C to 210 °C, demonstrating the good thermal stability. The corresponding fluctuation of $\Delta W_r$ and $\Delta \eta$ is calculated respectively to be less than 4.6 % and 3.2%, indicating the high-temperature application foreground. Secondly, fatigue-dependent *P-E* hysteresis loops were measured up to 10000 electric cycles at room temperature and the values of $W_r$ and $\eta$ were extracted, as shown in Figure 8 (b). A slight fluctuation of $\Delta W_r$<2.9% and $\Delta \eta$<3.1% is obtained, demonstrating the good fatigue stability. Lastly, the electric-field dependence of discharging performance was measured at room temperature, as shown in Figure 8 (c). The discharged current rapidly reaches a peak and releases completely within 2 μs. The peak current increases with the increment of electric field. From these curves, the discharged energy density $W_d$ as a function of time was calculated, as presented in inset of Figure 8 (c). The discharge time $\tau$, *i.e.*, the continuous time when 90% of the stored energy is released, is determined to be about 0.82 μs in the regime of applied electric fields.

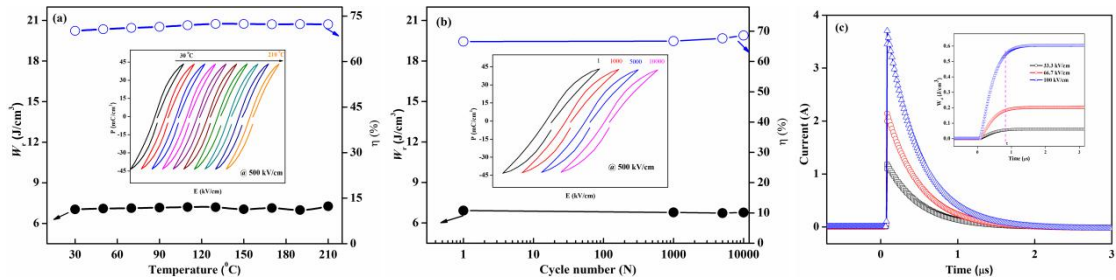

*Figure 8 Energy storage properties for NNO-0.1BHO thin film, (a) $W_r$ and $\eta$ values as a function of temperature, (b) $W_r$ and $\eta$ values as a function of cycle number, (c) discharging waveforms as a function of time at different electric field. Insets in (a) and (b) show the corresponding P-E hysteresis loops. Inset in (c) shows the calculated $W_d$ values*

## 4. Conclusion

In summary, a series of new (1-$x$)NaNbO$_3$-$x$BaHfO$_3$ ($x$=0, 0.05, 0.1, 0.15) lead-free thin films were grown on(001) Nb-doped SrTiO$_3$ single crystal substrates by a sol-gel method. X-ray diffraction results revealed the highly-preferred (00$l$)-orientation for all the thin films. The dense microstructure and the decreased grain size with increment of $x$ were demonstrated by atomic force microscopy. The addition of BaHfO$_3$ into NaNbO$_3$ improves significantly the leakage current and breakdown strength of pristine NaNbO$_3$ thin films, leading to the optimized energy storage performances. The superior recoverable energy storage density $W_r$ of 23.1 J/cm$^3$ at 1100 kV/cm was obtained at $x$=0.1. Also, the excellent thermal stability from 30 to 210°C ($\Delta W_r$ < 4.6%), the good fatigue resistance ($\Delta W_r$ < 2.9% after 10$^4$ electrical cycles), and the fast charge-discharge rate ($\tau$ = 0.82 μs ) were demonstrated for the optimized sample. These properties ascertain the great potential of environment-friendly (1-$x$)NaNbO$_3$-$x$BaHfO$_3$ thin films for energy storage applications in microelectronic devices.

**Acknowledgments**

This work is partially supported by the Fundamental Research Funds for the Central Universities (No: 310201911cx024) and Natural Science foundation of Shaanxi Province (No: 2021JM-059).

Molina-Luna, J. Koruza, Electric-field-induced antiferroelectric to ferroelectric phase transition in polycrystalline $NaNbO_3$, Acta Mater. 200 (2020) 127-135.

# Figure captions

Figure 1 (a) XRD patterns of NNO-$x$BHO thin films. Symbol (*) represents the secondary phase. (b) Phi-scan patterns of $x$=0.1 thin films deposited on NSTO substrates.

Figure 2 AFM images of NNO-$x$BHO thin films.

Figure 3 Current density ($J$) vs. electric field ($E$) curves for NNO-$x$BHO thin films.

Figure 4 Frequency-dependent dielectric constant (a) and dielectric loss (b) of NNO-$x$BHO thin films.

Figure 5 The Weibull distribution and the corresponding fitting results for NNO-$x$BHO thin films.

Figure 6 *P-E* hysteresis loops for NNO-$x$BHO thin films under different electric fields.

Figure 7 $W_r$ and $\eta$ values of NNO-$x$BHO thin films as a function of electric fields. Inset shows an enlarged view of $W_r$ curves for $x$=0 and $x$=0.1 thin films.

Figure 8 Energy storage properties for NNO-0.1BHO thin film, (a) $W_r$ and $\eta$ values as a function of temperature, (b) $W_r$ and $\eta$ values as a function of cycle number, (c) discharging waveforms as a function of time at different electric field. Insets in (a) and (b) show the corresponding *P-E* hysteresis loops. Inset in (c) shows the calculated $W_d$ values.